\documentclass[12pt]{spieman}  % 12pt font required by SPIE;
\usepackage{amsmath,amsfonts,amssymb}
\usepackage{graphicx}
\usepackage{setspace}
\usepackage{tocloft}

\title{Searching for missing interstellar oxygen in the far-infrared}

\author[a, *]{Takashi Onaka}
\author[b]{Itsuki Sakon}
\author[c]{Takashi Shimonishi}
\author[d]{Mitsuhiko Honda}
\affil[a]{The University of Tokyo, Graduate School of Science, Department of Astronomy, BUnkyo-ku, Tokyo 113-0033, Japan}
\affil[b]{The University of Tokyo, Graduate School of Science, Institute of Astronomy, Mitaka, Tokyo 181-0015, Japan}
\affil[c]{Niigata University, Institute of Science and Technology, Nishi-ku, Niigata 950-2181, Japan}
\affil[d]{Okayama University of Science, Department of Biosphere-Geosphere Science,Okayama, Okayama 700-0005, Japan}

\newcommand{\TO}[1]{\textcolor{black}{#1}}
\cftpagenumbersoff{figure}
\cftpagenumbersoff{table} 
\begin{document} 
\maketitle

\begin{abstract}
Study of interstellar elemental depletion poses an important problem in the interstellar matter that at least a quarter of the total oxygen ($\sim 160$\,ppm relative to hydrogen) is not accounted for in any known form of oxygen in the translucent or dense interstellar medium (ISM).  Detailed analysis of the absorption feature of water ice at 3\,$\mu$m suggests that one fifth of the missing oxygen may reside in 3\,$\mu$m-sized water ice grains.  However, the 3\,$\mu$m feature becomes complex and weak for grains larger than 3\,$\mu$m, and thus the NIR spectroscopy is not the best means to study the presence of large ice grains reliably.  Here we show that sensitive observations of the far-infrared (FIR) features of water ice at 44 and 62\,$\mu$m enable us to constrain the amount of crystalline water ice grains up to 5\,$\mu$m or even larger sizes unambiguously.  Oxygen is one of the key elements in the ISM chemistry, and [O I] 63\,$\mu$m is a dominant cooling line in the neutral ISM.  \TO{Understanding} the actual form of the missing oxygen in the ISM is crucial for the study of the ISM and star-formation process.  To detect the FIR features of the crystalline water ice over the expected strong continuum, a sensitive FIR spectrograph represented by PRIMA/FIRESS is indispensable.  Since the feature is broad, the low spectral resolution of $R \sim 130$ is sufficient, but accurate relative calibration better than 1\% is required.
\end{abstract}

% Include a list of up to six keywords after the abstract
\keywords{PRIMA; interstellar dust; elemental abundance; interstellar depletion; water ice; far-infrared spectroscopy }

% Include email contact information for corresponding author
{\noindent \footnotesize\textbf{*}Address all correspondence to Takashi Onaka,  \linkable{onaka@astron.s.u-tokyo.ac.jp} }

\begin{spacing}{1}   % use double spacing for rest of manuscript

\section{Introduction}
\label{sect:intro}  % \label{} allows reference to this section
Dust grains play an important role in the star-formation process of galaxies and control their evolution.  The actual compositions of dust grains, which are crucial for accurate understanding of their roles in the interstellar medium (ISM), however, remain an unresolved issue\cite{draine2003}.
The composition of interstellar dust is estimated in various ways.  Measurements
of the characteristic band features of individual solids from ultraviolet (UV) to infrared (IR) wavelengths
provide direct information
on the identification of the material, but the small particle effect originating
from the surface resonance and a combination of scattering and absorption effect of grains
whose size is comparable to the wavelength of the band in question
complicate unambiguous identification and accurate determination of compositions of the interstellar dust \cite{bohren1983, draine2003, henning2010}.  Elemental depletion measurements, an estimate
of the `depleted' amount of elements in the gas phase of the ISM relative to the reference abundance, 
on the other hand, give indirect,
but quantitative information on the elemental composition of dust grains, provided
that the deleted amount of the element is trapped in the solid phase\cite{jenkins2009}.  They also
give a crucial constraint on the dust model since the amount of the elements required by the model
must not exceed the observed amount of depletion.\TO{\cite{Mathis1996, Li1997}}  The interpretation of the depletion measurements
certainly depends on the `reference abundance', which is assumed as the total abundance of a given element both in the gas and solid phases.
In the past, the uncertainty in the reference abundance, which originated in
an apparent difference between the solar and B-star abundances, raised an issue of the
shortage of carbon in dust models \cite{snow1995, snow1996}.   \TO{Recent 
improvements in the analysis of both abundances mitigate the difference\cite{asplund2009, nieva2012}, while the protostellar abundance augmented by Galactic chemical enrichment is indicated to be an appropriate representqation of the interstellar abundance\cite{Zuo2021}}.

Together with the improved reference abundance, a detailed study of depletion observations poses a new, even more significant issue, missing oxygen or unidentified depleted oxygen (UDO)
in the ISM.\TO{\cite{whittet2010, Wang2015}}  Ref.~\citenum{jenkins2009} first points out that about a quarter of the total
oxygen ($\sim 160$\,ppm relative to hydrogen) is missing in its manifestation in diffuse interstellar clouds.  Since
oxygen generally becomes solid as silicates or metal oxides, it requires metal elements to be combined into solid phase.
However, the required metal abundance to account for the missing oxygen 
far exceeds the reference abundance and thus
the missing amount cannot be attributed to any silicate or metal oxide grains. While recent X-ray spectroscopy suggests that there may not be missing oxygen in the less dense, diffuse ISM\cite{Psaradaki2023, Psaradaki2024},
extrapolation of the trend with the gas density into denser regions, where part of
oxygen also is incorporated into CO gas and ices, indicates that the amount of the missing oxygen could be even larger ($\sim 300$\,ppm)
in dense ($\sim 1000$\,cm$^{-3}$) regions of the ISM.\cite{whittet2010}
To explain the missing oxygen, several possibilities have been discussed.
While it may be sequestered into an organic carbonate\cite{Jones2019}, it is also suggested
that a large fraction of the missing oxygen could be hiding in large ice grains since water ice
is only oxygen-bearing species that can be in the solid phase without conflicting the abundance constraint.\cite{jenkins2009}
NIR spectroscopy \TO{using the} Infrared Camera on board AKARI reveals the ubiquitous presence of the broad, shallow 3\,$\mu$m absorption feature of water ice in star-forming regions of our Galaxy,\cite{mori2014,onaka2014, onaka2022} supporting the possibility that the missing oxygen may be sequestered in large water ice grains.

In fact, micron-sized water ice grains show complicated structures at the strong 3\,$\mu$m absorption
band due to the effect of scattering\cite{Dartois2024} and thus can easily be elusive in the 3\,$\mu$m spectroscopy.  
This possibility is investigated in detail for the well-studied line of sight to $\zeta$Oph
based on the 3\,$\mu$m spectrum taken with {\it ISO}-SWS, and tentative evidence for the presence
of $2.8$\,$\mu$m size ice grains is suggested.\cite{poteet2015}  However, even including possible amounts of Fe$_3$O$_4$ dust
and polyoxymethylene (POM), they can account only for one third ($\sim 37$\%) of the missing oxygen.
Since the 3\,$\mu$m absorption of water ice becomes weak and complicated
as the particle size increases, it is rather difficult to study the presence of 
ice grains lager than 3\,$\mu$m unambiguously.  The 3\,$\mu$m spectroscopy may not be the best means to study their presence.  
Alternatively, high energy-resolution spectroscopy at
the oxygen K-edge (540\,eV) in the X-ray offers
an independent way to study the nature of oxygen-bearing species in both the gas and solid phases in the ISM\cite{paerels2001, takei2002, constantini2005, baumgartner2006, lee2009, Psaradaki2023}.
However, large grains ($> 1$\,$\mu$m) become optically thick at the oxygen K-edge  and
thus the X-ray spectroscopy is not sensitive to \TO{the presence of oxygen}.
Far-infrared (FIR) features do not have this problem unless grains are too large, e.g., 
larger than 10\,$\mu$m, and thus offer a unique means to study the presence of large
ice grains.  Taking account of this advantage, we investigate
the possibility to constrain the fraction of the missing oxygen hiding in large
water ice grains.

\section{Infrared Properties of Water Ice}
\label{sect:IRproperties}
Water ice has various forms in crystalline and amorphous phases.  The crystalline structure below 140\,K is
the cubic form $I_\mathrm{c}$, while at higher temperatures it becomes the hexagonal from $I_\mathrm{h}$.  The spectra in
30--200\,$\mu$m of both crystalline forms are indistinguishable \cite{bertie1967}.  The FIR spectrum of
$I_\mathrm{h}$ shows distinct lattice vibration bands at 44, 52, and 62\,$\mu$m, which are attributed
to the transverse optical (TO), longitudinal optical (LO), and longitudinal acoustic (LA) branches, respectively
 \cite{bertie1967}.  Among them, the 44\,$\mu$m band is the strongest.  Amorphous water ice has two forms;
 high-density (1.1\,g\,cm$^{-3}$) and low-density (0.94\,g\,cm$^{-3}$) ones.  The change from the former
 to the latter phase occurs in an irreversible manner between 38 and 80\,K \cite{jenniskens1995}.
Amorphous water ice shows a broad band feature at around 45\,$\mu$m \cite{hudgins1993}.
 
 In the following analysis, we adopt the refractive index provided
 by Ref.~\citenum{bertie1969} for $I_\mathrm{h}$ crystalline ice and that for amorphous ice by Ref.~\citenum{hudgins1993} at 10\,K for 2.5--200\,$\mu$m.
 The phase change from amorphous to crystalline due to the increase of temperature and the irradiation of
 protons has been studied extensively\cite{moore1992, smith1994}. Changes 
associated with the phase change are seen in the FIR spectrum. 
On the other hand, the change within a given phase is small and not crucial for the following discussion.\cite{smith1994}  Therefore, we do not take account of the spectral change with the temperature nor thermal history of amorphous ice.
In interstellar clouds, where ice grains efficiently form, coagulation
may also occur in addition to the accretion of ice onto pre-existing silicate or carbon grains, and
thus silicate or carbon grains could be distributed randomly in ice grains.  
In the following, we assume that ice grains are \TO{spherical with radius $a$} and 
contain a small volume fraction of carbon or silicate inclusions. 
We
employ the effective medium theory of Maxwell Garnett for inhomogeneous medium to calculate the average
refractive index\cite{bohren1983}, which is different from the calculation of core-mantle grains in the previous study\cite{poteet2015}.   The difference between the two calculations is negligible as far as the volume fraction
of the inclusions is small and does not affect the conclusion of this paper.  
For the refractive index, we use that of astronomical silicate \cite{draine1985} for silicate
and that of BE carbon \cite{zubko1996} for carbon grains.  For the refractive index of ice for both crystalline and amorphous forms
in the UV to optical region, we use that of $I_\mathrm{h}$ ice provided by Ref.~\citenum{warren2008} simply because
no data are available for amorphous ice.  Since we do not discuss spectral features in the UV to optical
regions, this does not affect the following analysis.
In the following, we set the volume fraction of carbon or silicate inclusions as 0.1, following the previous study, which assumed it as 0.125\cite{poteet2015}.  We calculate the absorption and scattering cross-sections using the Mie theory\cite{bohren1983}.

\begin{figure}
\begin{center}
\includegraphics[width=40pc]{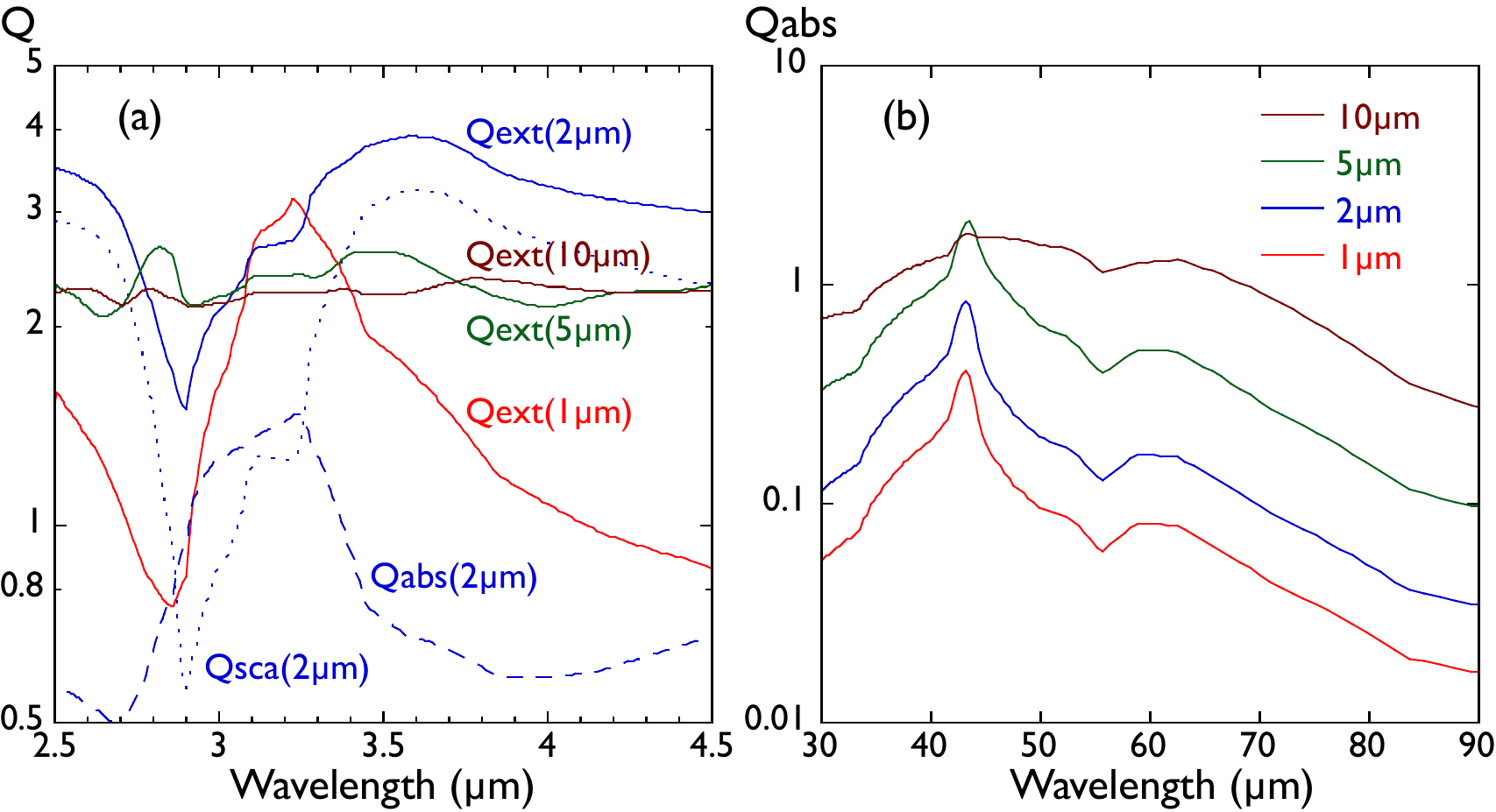}
\caption{Efficiency factors of crystalline ice grains with carbon inclusions for grain radii of 1 (red), 2 (blue), 5 (green), and 10\,$\mu$m (brown).
The inclusion fraction is set as 0.1.
(a) The solid lines indicate the extinction efficiency factors \TO{in the 3\,$\mu$m region}.
The scattering and absorption efficiency factors for a 2\,$\mu$m grain are also shown with the dotted and dashed lines.
(b) Absorption efficiency factors in the FIR.  The colors are the same as in (a).}\label{fig_ice1}
\end{center}
\end{figure}

\begin{figure}
\begin{center}
\includegraphics[width=40pc]{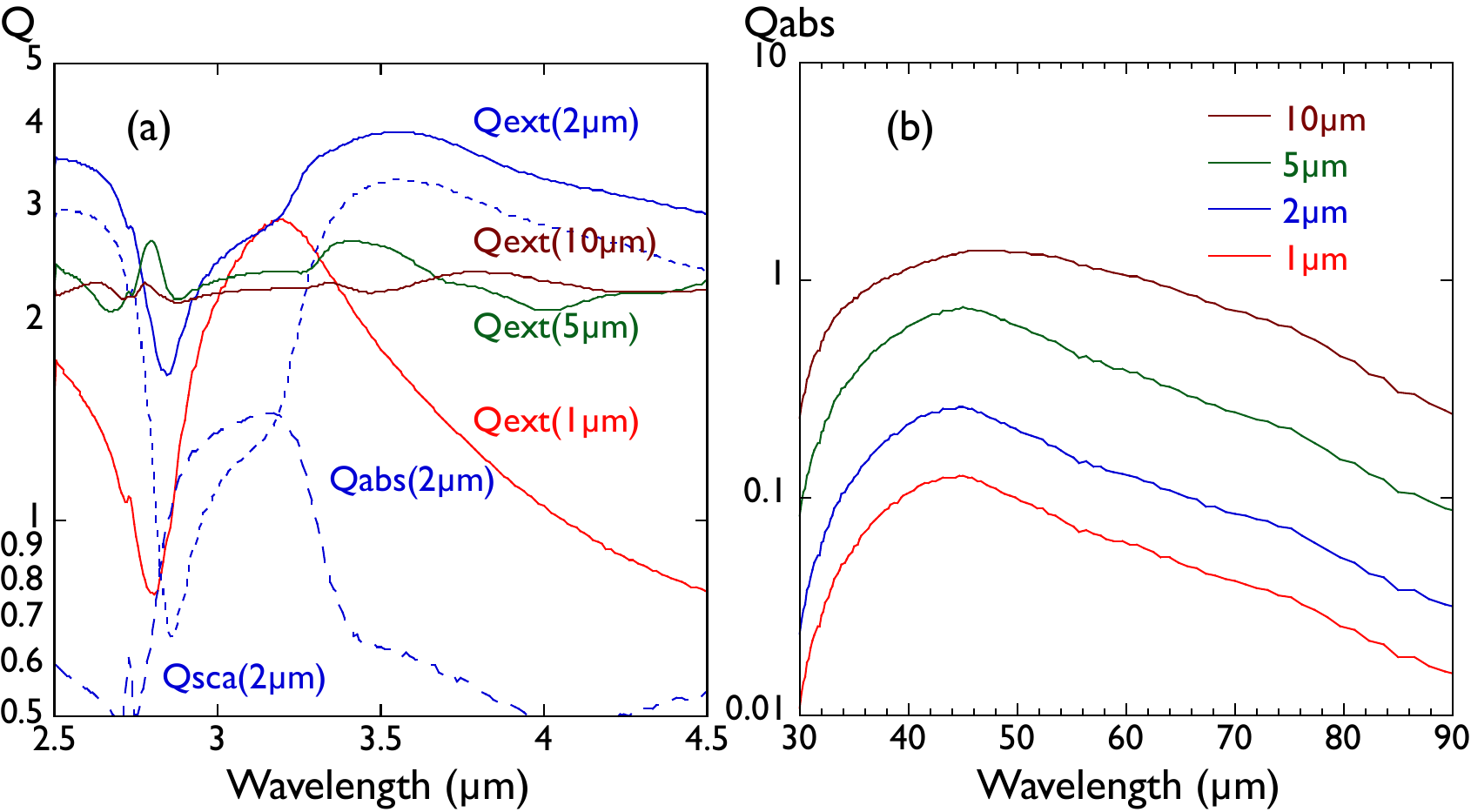}
\caption{Same as Fig.~\ref{fig_ice1} except for amorphous ice grains.}\label{fig_ice2}
\end{center}
\end{figure}

Figure~\ref{fig_ice1}a shows the efficiency factors (cross-section divided by $\pi a^2$) of crystalline ice grains with carbon inclusions
for radii of 1, 2, 5, and 10\,$\mu$m at
2.5--4.5\,$\mu$m, \TO{and} Figure~\ref{fig_ice1}b indicates the absorption efficiency factors in the FIR.
Ice grains with silicate inclusions show very similar results except for a slight difference in the strength of the underlying continuum.
The extinction efficiency factor $Q_\mathrm{ext}$ becomes broad and complex for a grain with $a=2$\,$\mu$m as
the scattering ($Q_\mathrm{sca}$) becomes dominant in the longer wavelength side (blue solid and dotted lines in Fig.~\ref{fig_ice1}).  This confirms
the results of previous calculations of core-mantle grains with radii of 2.0--3.2\,$\mu$m.\cite{poteet2015}
Here, we expand the radius range and show that the band structure in the extinction efficiency
becomes very small for grains with $a > 5$\,$\mu$m.  The peak of the extinction efficiency factor in the longer wavelength, which comes from the scattering, shifts with the grains size.  If there is a size distribution of ice grains, therefore, the band feature will further be smoothed out\cite{poteet2015}, suggesting that ice grains larger than 3\,$\mu$m are
difficult to be detected reliably with the 3\,$\mu$m spectroscopy.  In the FIR (Fig.~\ref{fig_ice1}b), on the other hand, the 44 and
62\,$\mu$m bands do not change their profiles and are proportional to the volume of the grain up to a grain size of $\sim 5$\,$\mu$m.  Hence, the amount of oxygen atoms in the water ice grains can be estimated accurately up to a $\sim 5$\,$\mu$m size.  The FIR features are
recognizable even for grains with $a =10$\,$\mu$m.  Figure~\ref{fig_ice2} shows the results for
amorphous ice grains.  A similar trend to Fig.~\ref{fig_ice1} is seen in the 3\,$\mu$m region.  In the FIR
(Fig.~\ref{fig_ice2}b), the band is very broad, and it is not easy to  distinguish it from the underlying continuum
even for 1\,$\mu$m-sized grains.

\section{Detectability of Large Ice Grains in the FIR}
\label{sect:detectability}
\begin{figure}
\begin{center}
\includegraphics[width=20pc]{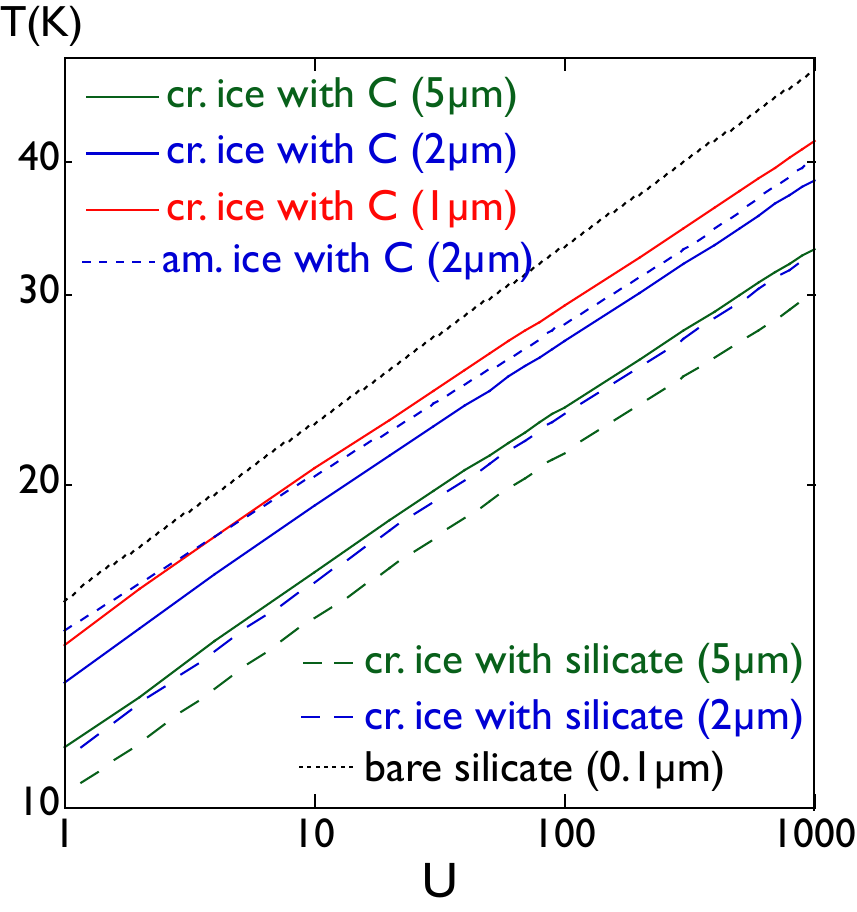}
\caption{Radiative equilibrium temperatures of various grains against the scaled ISRF intensity $U$.
\TO{The red, blue, and green solid lines indicate the temperatures of crystalline ice grains of radii of 1, 2, and 5\,$\mu$m with carbon inclusions, respectively, and
the blue and green long dashed lines show the temperatures of crystalline ice grains of radii of 2 and 5\,$\mu$m with silicate inclusions, respectively.  The blue short dashed line indicates the temperature of an amorphous ice grain of a radius of 2\,$\mu$m with carbon inclusions.}
The black dotted line indicates the temperature of a bare silicate grain of a radius of 0.1\,$\mu$m as a representative
of the underlying continuum of the thermal emission of the diffuse ISM.  \TO{The number in the parentheses in the legend indicates the size of the grain.}}\label{fig_ice3}
\end{center}
\end{figure}

In this section, we investigate the detectability of the FIR features in emission spectra of the ISM.
We use the DUSTEM model \cite{compiegne2011} to calculate the underlying continuum emission from
silicate and carbonaceous grains and
add the emission from ice grains onto it.  We assume the standard parameters in the DUSTEM
model and vary the strength of the interstellar radiation field (ISRF) $U$, where $U$ is a scaling factor relative to the ISRF of solar neighborhood\cite{Mathis1983}.  Figure~\ref{fig_ice3} shows
the expected radiative equilibrium temperatures for ice grains with various radii and a silicate grain
of 0.1\,$\mu$m radius.  The latter is regarded as a representative of the temperature of the underlying continuum.  \TO{The radii of the grains are indicated in the parentheses in the legend.}

Figure~\ref{fig_ice3} shows that a silicate grain of 0.1\,$\mu$m radius has a higher temperature
than large ice grains.  This is partly due to the grain size effect since larger grains 
have larger emissivity in the IR and partly to the
strong FIR band of crystalline ice at 44\,$\mu$m.  Amorphous ice grains have higher temperatures than 
crystalline ice grains because their FIR band is weaker.
Ice grains with carbon inclusions have higher temperatures than those with
silicate inclusions because of higher absorptivity of carbon inclusions in the UV to optical regions.

\begin{figure*}
\begin{center}
\includegraphics[width=40pc]{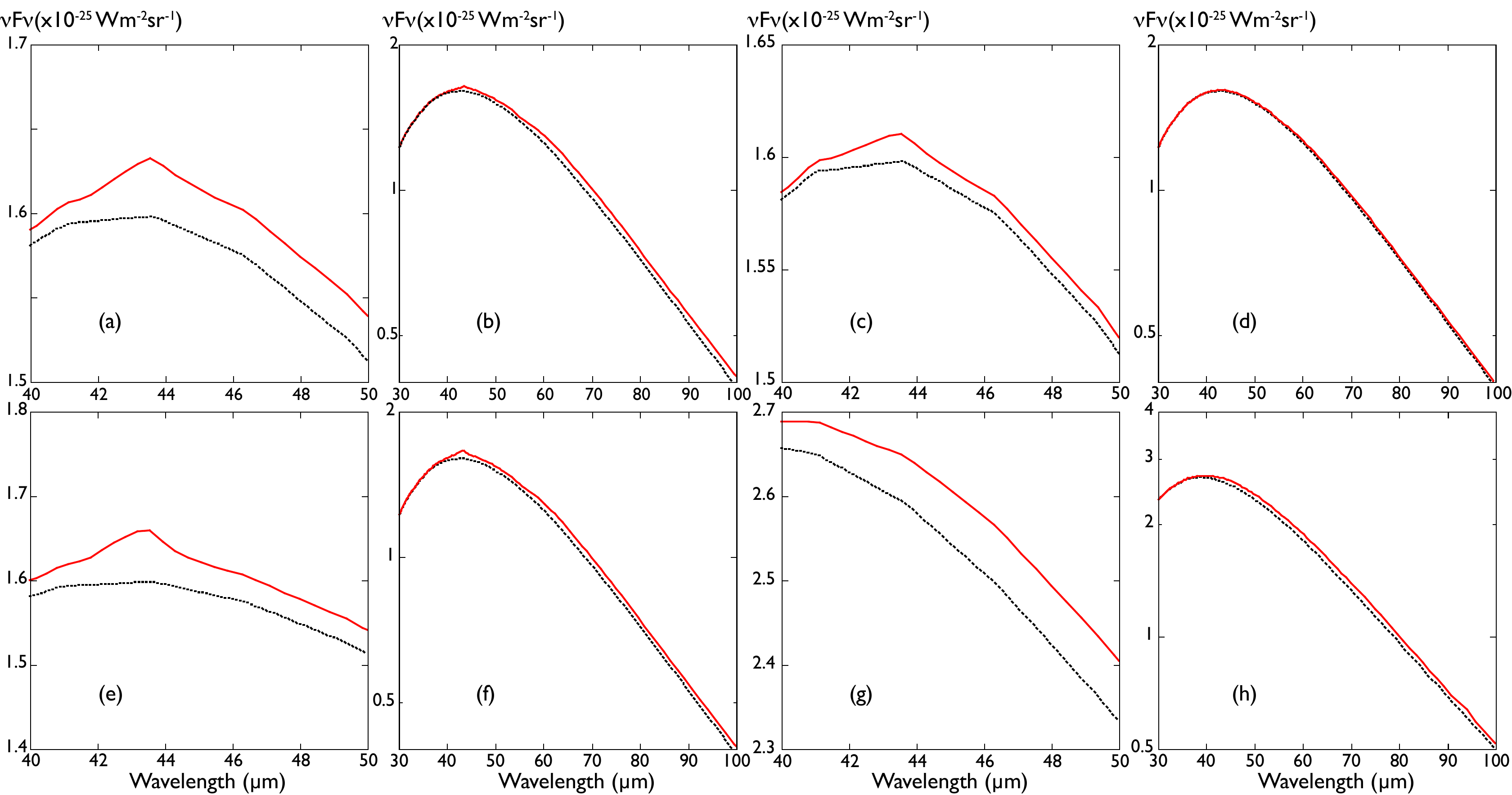}
\caption{Model SEDs of the emission from the ISM without ice grains (black dotted lines) and with ice grains
(red solid lines). Units are per hydrogen atom. (a) and (b) show the SEDs with crystalline ice grains with carbon inclusions of the 5\,$\mu$m radius for the ISRF
intensity of $U = 600$, which corresponds to $T_\mathrm{ice}= 30.8$\,K.  (c) and (d) indicate the SEDs with crystalline
ice grains with silicate inclusions of the 2\,$\mu$m radius in $U=600$ and $T_\mathrm{ice}=30.4$\,K.  (e) and (f) show the SEDs with crystalline
ice grains with carbon inclusions of the 2\,$\mu$m radius for $U=600$ and $T_\mathrm{ice}=35.6$\,K.  (g) and (h) indicate the SEDs
with amorphous ice grains with carbon inclusions of the 2\,$\mu$m radius for $U=1000$ and $T_\mathrm{ice}=40.1$\,K. (a), (c), (e), and (g) are enlarged versions for 40--50\,$\mu$m of (b), (d), (f), and (h), respectively.}\label{fig_ice4}
\end{center}
\end{figure*}

Figure~\ref{fig_ice4} plots the model spectral energy distributions (SEDs) of the diffuse IR emission with various ice grains.
The parameters for the SED calculations are (1) crystalline (cr) or amorphous (am) ice phase, (2) grain radius $a$, (3) 
carbon or silicate inclusions, and the intensity of the ISRF $U$.  The ice grain temperature
$T_\mathrm{ice}$ is calculated for a given $U$ (Fig.~\ref{fig_ice3}).  The underlying continua are calculated by the DUSTEM
model with the standard input parameters for various $U$'s. 
We simply add emission from ice grains in a given interstellar radiation intensity $U$, assuming that the oxygen
abundance in water ice is 160\,ppm, corresponding to the amount of the UDO in the diffuse ISM \cite{poteet2015}, to the SEDs
from the DUSTEM model.

Figure~\ref{fig_ice4} suggests that for crystalline ice grains,
the FIR band at 44\,$\mu$m is discernible if $T_\mathrm{ice}>30$\,K, where $T_\mathrm{ice}$ is the temperature of the ice grain, and if the observed spectrum has a sufficiently
high signal-to-noise ratio (SNR).  To make it more quantitatively, we calculate the emission at 44\,$\mu$m over the linearly
interpolated spectrum between 40 and 47\,$\mu$m and compare it for the SEDs with and without ice
emission.  We found that the difference is about 1\% for the case of carbon inclusions, $a=5$\,$\mu$m, and $U=600$
($T_\mathrm{ice}=30.8$\,K, Fig.~\ref{fig_ice4}a).  In this case, the excess at the band can barely be seen in the spectrum in a wider
spectral range (Fig.~\ref{fig_ice4}b), if the observed spectrum has a SNR larger than 100.
For the case of silicate inclusions, $a=2$\,$\mu$m and $U=600$ ($T_\mathrm{ice}=30.4$\,K), 
however, the difference is only 0.4\% and the band feature is not easily recognized in the spectrum in a wider spectral range
(Fig.~\ref{fig_ice4}d).  It requires SNR larger than 200 to detect the FIR feature.
The situation is improved if $T_\mathrm{ice}$ becomes higher than 35\,K (Figs.~\ref{fig_ice4}e and f), where the difference becomes
2\%.  The reliability of the detection thus  increases with the ice temperature significantly.  
Figures~\ref{fig_ice4}g and h suggest that the FIR feature of amorphous ice at around 45\,$\mu$m cannot be clearly seen 
even in a very high radiation field environment ($U=1000$ and $T_\mathrm{ice}=40.1$\,K)~\cite{Kamp2018, Kamp2021}.
It is almost impossible to distinguish a very broad feature expected for amorphous ice at around 45\,$\mu$m
unless we have very accurate knowledge of the underlying continuum.    

We conclude from these simple model calculations that if a certain fraction of ice is in the crystalline form and the temperature is higher than 30\,K,
we are able to detect the FIR feature at 44\,$\mu$m.   Temperatures higher than 35\,K will lead to reliable detection.  
These conditions correspond to
$U$ larger than 600 for the grain size of 2--5\,$\mu$m in the present calculation.  It should be noted that the ice temperature 
also depends on the volume fraction of the inclusions.  
%For lower fraction of inclusions, the expected temperature decreases and
%higher radiation field intensities are required for detection.  For instance, if the volume fraction of carbon inclusions 
%becomes only 1\%, the expected temperature for a 2\,$\mu$m ice grain is 30.9\,K at $U=500$, whereas the temperature of
%a 2\,$\mu$m ice grain with carbon inclusions of 10\% is 30.2\,K at $U=200$.
If the volume fraction is 0.05, then the expected temperature at $U=200$ becomes 29.4\,K, compared to 30.2\,K for the volume fraction of 0.1.

Figure~\ref{fig_ice4} shows that the peak target flux will be $1.6 \times 10^{-25}$\,W\,m$^{-2}$\,sr$^{-1}$ per hydrogen. Assuming a hydrogen column density of $10^{20}$\,cm$^{-2}$, the predicted surface brightness will be $1.6 \times 10^{-5}$\,W\,m$^{-2}$\,sr$^{-1}$at 40\,$\mu$m. 
Low-resolution of $R\sim 130$ would be sufficient for the
detection of the water ice features, but it would require high
SNRs, more than 100.  High sensitivity of PRIMA/FIRESS will achieve these requirements within a reasonable integration time with the small mapping mode (see Bradford et al. in this volume).  Highly accurate calibration will be a more important requirement for this study.  Because of the strong continuum expected underneath, the relative calibration must be better than 1\%.

\section{Discussion}
\label{section:discussion}
Excess extinction at 4--8\,$\mu$m over the model prediction is reported toward the Galactic center based on {\it ISO} SWS observations.\cite{lutz1996}.  Observations with {\it Spitzer} confirm the flat extinction in this spectral range\cite{indebetouw2005, flaherty2007, wang2013}, while recent studies indicate that the extinction in the NIR varies \TO{significantly}, depending on the line-of-sight\cite{Declei2022, Gordon2023}. The flat extinction can be interpreted in terms of the presence of micron-sized grains \cite{wang2014, Wang2015}.  
The extinction from the near- to mid-infrared (1--25\,$\mu$m) in dense clouds also suggests the presence of large grains, although there is no evidence for
the progressive grain growth in dense clouds for $A_\mathrm{K} > 1$ \cite{boogert2011}.  Observations of 'coreshine'
indicate the presence of micron-sized grains in some cores.\cite{steinacker2015, Ysard2016, Saajasto2018}  The so-called `red wing', excess absorption seen in 3.3--3.7\,$\mu$m, in the water ice absorption feature at 3\,$\mu$m
can partly be attributed to large ice grains.\cite{boogert2015}
Recent JWST observations also clearly show the presence of micron-sized ice grains in the dense Chamaeleon I cloud\cite{Dartois2024}.
It should also be noted that micron-sized grains of interstellar origin were captured by the Stardust Interstellar Dust Collector\cite{westphal2014}.  
Therefore, the presence of micron-sized grains in the ISM \TO{is confirmed} by observations as well as collected samples {\it in situ}.  \TO{It is also shown that since large ice grains have  flat extinction in the UV to optical and lower temperatures than bare silicate and carbonaceous grains, they do not make significant effects on the UV/optical extinction curve and the FIR emission.\cite{Wang2015}}

The 3\,$\mu$m ice absorption feature of several objects also suggests the presence of crystalline ice grains.\cite{boogert2015}
About 10\% of ice needs to be crystalline to account for the
3\,$\mu$m feature seen in the Becklin-Neugebauer object,\cite{smith1989} while the observed 3\,$\mu$m absorption profile suggests more than 70\% of ice in the crystalline form in the silhouette disk d216-0939.\cite{terada2012}  
Detection of
the FIR emission features at 44 and 62\,$\mu$m of crystalline ice 
is reported in \TO{several OH/IR stars, including} IRAS 09371+1212,\cite{omont1990}
where crystalline ice is expected to be formed directly from high-temperature gas in the stellar outflow.  \TO{The Infrared Space Observatory (ISO) further detects these FIR ice features in several more objects.\cite{Sylvester1999}}
The 62\,$\mu$m crystalline ice band is detected in the
Herbig-Haro object HH7, where it is suggested to be formed in shocks.\cite{molinari1999}
The 44\,$\mu$m band is also seen in absorption toward two embedded IR sources.\cite{dartois1998}   The presence of
the 44 and 62\,$\mu$m emission features is indicated toward the disk of the Herbig star HD 142527\cite{min2016} and the 62\,$\mu$m is detected in emission for several more sources in planet-forming regions.\cite{McClure2012,McClure2015}
Except for the case of \TO{OH/IR stars},  crystalline ice is thought to be formed from crystallization of amorphous ice
by thermal processing, e.g., shocks or radiation.  Possible mechanisms of formation
of crystalline water ice in the ISM have been discussed in detail, including brief heating of grains in grain-grain collisions or cosmic-ray impacts.\cite{jenniskens1996}
Amorphous ice under UV irradiation behaves like liquids at 65--150K.\cite{tachibana2017}
The `liquid-like' amorphous ice has measurable viscosity, suggesting that crystallization of amorphous water ice may proceed in a finite time in diffuse clouds, 
where the UV radiation
intensity is relatively modest.\cite{tanaka2019}
These studies suggest a possibility that an appreciable part of water ice in the ISM is in crystalline form, although the exact fraction is difficult to be estimated at the present.

\TO{Photodesorption or photosputtering by the UV radiation is  crucial for the survival of water ice grains in the ISM.\cite{Wang2015}  Laboratory data of the photodesoprtion yield of water ice have some uncertainty.\cite{Cuppen2024}  Recent data\cite{Cruz-Diaz2018} show slightly smaller values with less temperature dependence than previous ones,\cite{Oberg2009} which
suggest that the survival time scale of water ice grains of a radius of 5\,$\mu$m in the ISM is more than $9.2 \times 10^6$\,yr at temperatures of $8-50$\,K.  The survival time scale will be even longer for grains larger than 5\,$\mu$m. Since the diffuse ISM is supposed to be replenished by matters from molecular clouds with a time scale of $\sim 3 \times 10^6 - 2 \times 10^7$\,yr,\cite{Draine1990} it seems likely that micron-sized water ice grains survive in the diffuse ISM long enough to be replenished by the mass supply from molecular clouds, while submicron-sized ice grains vanish quickly.\cite{Wang2015}}

\section{Conclusion}
\label{section:conclusion}
Using the simple dust emission model, the present study shows that FIR spectroscopy of the ISM provides us with unique information
on the amount of crystalline ice grains in the ISM, even for grains larger than 3\,$\mu$m, which cannot be studied by other means, if the ice temperature is sufficiently high ($> 35$\,K).  
If we can make observations toward objects where relatively high ice temperatures are expected, we should be able to constrain the amount of ice grains up to $\sim 10$\,$\mu$m size reliably. 
The exact temperature of ice grains depends on various factors, e.g., the size of the grain, and the fraction and properties of the inclusions.
For the size range of 2--5\,$\mu$m, the strengths of FIR features are still proportional to
the volume of ice grains and thus the amount of ice can be estimated once the temperature of the ice grain is accurately determined.
The temperature of large ice grains could be different from small silicate and carbon grains, and thus we 
have to estimate it independently.
If we could detect the other feature of ice at around 62\,$\mu$m, we can
have a reliable estimate of the temperature.  Alternatively, we may also use the \TO{dependence} of the peak position of the 44\,$\mu$m feature
on the temperature.\cite{moore1992} 
While there are no known strong gas emission lines nor strong dust features at around 44\,$\mu$m, detection of another 62\,$\mu$m band will certainly secure the detection of the water ice grain, since the feature is expected to be faint.

Oxygen is the third most abundant element in the Universe.  It is an important element in the terrestrial atmosphere and for the presence of life.  Therefore, its actual form in the ISM has significant implications on our understanding of the interstellar physics as well as the origin of life.
In the present analysis, we fix the oxygen abundance as 160\,ppm.  
If the grain temperature is sufficiently high, we should be able to detect a fraction of this amount of ice grains, which will
provide useful information on the study of the missing oxygen in the ISM.  It should also be added that
[OI]\,63\,$\mu$m is an important coolant in dense PDRs as well as in slow shocks with medium to high density.\cite{draine1983, hollenbach1997, lesaffre2013}  \TO{In PDRs, even large water ice grains will be destroyed quickly by photodesorption because of the strong UV irradiation, which affects the oxygen abundance in the gas phase.}
It is thus crucial to estimate an accurate amount of oxygen in the gas phase for the study of the energy balance in the ISM.
Sensitive FIR spectroscopy of 40--200\,$\mu$m offers a unique means to study both solid and gaseous oxygen at the same time,  making
a significant contribution to our understanding of interstellar physics.

\subsection*{Disclosures}
The authors declare there are no financial interests, commercial affiliations, or other potential conflicts of interest that have influenced the objectivity of this research or the writing of this paper.

\subsection* {Code, Data, and Materials Availability} 
The materials utilized in this study can be requested from the author at \linkable{onaka@astron.s.u-tokyo.ac.jp}.

\section* {Acknowledgments}
The authors appreciate the large efforts of the editors who organize this special issue as well as the PRIMA project team.  TO thanks Bruce Draine for pointing out the issue of the UDO
and the paper by Ref.~\citenum{poteet2015}.
This work is supported in part by JSPS KAKENHI No. JP24K07087.

%%%%% References %%%%%

\bibliography{onaka}   % bibliography data in report.bib

\begin{thebibliography}{10}

\bibitem{draine2003}
B.~T. {Draine}, ``{Interstellar Dust Grains},'' {\em Annu. Rev. Astron.
  Astrophys.} {\bf 41}, 241--289  (2003).

\bibitem{bohren1983}
C.~F. {Bohren} and D.~R. {Huffman}, {\em {Absorption and scattering of light by
  small particles}}, Wiley, New York  (1983).

\bibitem{henning2010}
T.~{Henning}, ``{Cosmic Silicates},'' {\em Annu. Rev. Astron. Astrophys.} {\bf
  48}, 21--46  (2010).

\bibitem{jenkins2009}
E.~B. {Jenkins}, ``{A Unified Representation of Gas-Phase Element Depletions in
  the Interstellar Medium},'' {\em Astrophys. J.} {\bf 700}, 1299--1348
  (2009).

\bibitem{Mathis1996}
J.~S. {Mathis}, ``{Dust Models with Tight Abundance Constraints},'' {\em
  Astrophys. J.} {\bf 472}, 643  (1996).

\bibitem{Li1997}
A.~{Li} and J.~M. {Greenberg}, ``{A unified model of interstellar dust.},''
  {\em Astron. Astrophys.} {\bf 323}, 566--584  (1997).

\bibitem{snow1995}
T.~P. {Snow} and A.~N. {Witt}, ``{The Interstellar Carbon Budget and the Role
  of Carbon in Dust and Large Molecules},'' {\em Science} {\bf 270}, 1455--1460
   (1995).

\bibitem{snow1996}
T.~P. {Snow} and A.~N. {Witt}, ``{Interstellar Depletions Updated: Where All
  the Atoms Went},'' {\em Astrophys. J. Lett.} {\bf 468}, L65  (1996).

\bibitem{asplund2009}
M.~{Asplund}, N.~{Grevesse}, A.~J. {Sauval}, {\em et~al.}, ``{The Chemical
  Composition of the Sun},'' {\em Annu. Rev. Astron. Astrophys.} {\bf 47},
  481--522  (2009).

\bibitem{nieva2012}
M.-F. {Nieva} and N.~{Przybilla}, ``{Present-day cosmic abundances. A
  comprehensive study of nearby early B-type stars and implications for stellar
  and Galactic evolution and interstellar dust models},'' {\em Astron.
  Astrophys.} {\bf 539}, A143  (2012).

\bibitem{Zuo2021}
W.~{Zuo}, A.~{Li}, and G.~{Zhao}, ``{Interstellar Extinction and Elemental
  Abundances},'' {\em Astrophys. J. Suppl.} {\bf 252}, 22  (2021).

\bibitem{whittet2010}
D.~C.~B. {Whittet}, ``{Oxygen Depletion in the Interstellar Medium:
  Implications for Grain Models and the Distribution of Elemental Oxygen},''
  {\em Astrophys. J.} {\bf 710}, 1009--1016  (2010).

\bibitem{Wang2015}
S.~{Wang}, A.~{Li}, and B.~W. {Jiang}, ``{The interstellar oxygen crisis, or
  where have all the oxygen atoms gone?},'' {\em Mon. Not. R. Astron. Soc.}
  {\bf 454}, 569--575  (2015).

\bibitem{Psaradaki2023}
I.~{Psaradaki}, E.~{Costantini}, D.~{Rogantini}, {\em et~al.}, ``{Oxygen and
  iron in interstellar dust: An X-ray investigation},'' {\em Astron.
  Astrophys.} {\bf 670}, A30  (2023).

\bibitem{Psaradaki2024}
I.~{Psaradaki}, L.~{Corrales}, J.~{Werk}, {\em et~al.}, ``{Elemental Abundances
  in the Diffuse Interstellar Medium from Joint Far-ultraviolet and X-Ray
  Spectroscopy: Iron, Oxygen, Carbon, and Sulfur},'' {\em Astron. J.} {\bf
  167}, 217  (2024).

\bibitem{Jones2019}
A.~P. {Jones} and N.~{Ysard}, ``{The essential elements of dust evolution. A
  viable solution to the interstellar oxygen depletion problem?},'' {\em
  Astron. Astrophys.} {\bf 627}, A38  (2019).

\bibitem{mori2014}
T.~I. {Mori}, T.~{Onaka}, I.~{Sakon}, {\em et~al.}, ``{Observational Studies on
  the Near-infrared Unidentified Emission Bands in Galactic H II Regions},''
  {\em Astrophys. J.} {\bf 784}, 53  (2014).

\bibitem{onaka2014}
T.~{Onaka}, T.~I. {Mori}, I.~{Sakon}, {\em et~al.}, ``{Search for the Infrared
  Emission Features from Deuterated Interstellar Polycyclic Aromatic
  Hydrocarbons},'' {\em Astrophys. J.} {\bf 780}, 114  (2014).

\bibitem{onaka2022}
T.~{Onaka}, I.~{Sakon}, and T.~{Shimonishi}, ``{Near-infrared Spectroscopy of a
  Massive Young Stellar Object in the Direction toward the Galactic Center: XCN
  and Aromatic C-D Features},'' {\em Astrophys. J.} {\bf 941}, 190  (2022).

\bibitem{Dartois2024}
E.~{Dartois}, J.~A. {Noble}, P.~{Caselli}, {\em et~al.}, ``{Spectroscopic
  sizing of interstellar icy grains with JWST},'' {\em Nature Astronomy} {\bf
  8}, 359--367  (2024).

\bibitem{poteet2015}
C.~A. {Poteet}, D.~C.~B. {Whittet}, and B.~T. {Draine}, ``{The Composition of
  Interstellar Grains toward {$\zeta$} Ophiuchi: Constraining the Elemental
  Budget near the Diffuse-dense Cloud Transition},'' {\em Astrophys. J.} {\bf
  801}, 110  (2015).

\bibitem{paerels2001}
F.~{Paerels}, A.~C. {Brinkman}, R.~L.~J. {van der Meer}, {\em et~al.},
  ``{Interstellar X-Ray Absorption Spectroscopy of Oxygen, Neon, and Iron with
  the CHANDRA LETGS Spectrum of X0614+091},'' {\em Astrophys. J.} {\bf 546},
  338--344  (2001).

\bibitem{takei2002}
Y.~{Takei}, R.~{Fujimoto}, K.~{Mitsuda}, {\em et~al.}, ``{On and Ne K
  Absorption Edge Structures and Interstellar Abundance toward Cygnus X-2},''
  {\em Astrophys. J.} {\bf 581}, 307--314  (2002).

\bibitem{constantini2005}
E.~{Costantini}, M.~J. {Freyberg}, and P.~{Predehl}, ``{Absorption and
  scattering by interstellar dust: an XMM-Newton observation of Cyg X-2},''
  {\em Astron. Astrophys.} {\bf 444}, 187--200  (2005).

\bibitem{baumgartner2006}
W.~H. {Baumgartner} and R.~F. {Mushotzky}, ``{Oxygen Abundances in the Milky
  Way Using X-Ray Absorption Measurements toward Galaxy Clusters},'' {\em
  Astrophys. J.} {\bf 639}, 929--940  (2006).

\bibitem{lee2009}
J.~C. {Lee}, J.~{Xiang}, B.~{Ravel}, {\em et~al.}, ``{Condensed Matter
  Astrophysics: A Prescription for Determining the Species-specific Composition
  and Quantity of Interstellar Dust Using X-rays},'' {\em Astrophys. J.} {\bf
  702}, 970--979  (2009).

\bibitem{bertie1967}
J.~E. {Bertie} and E.~{Whalley}, ``{Optical Spectra of Orientationally
  Disordered Crystals. II. Infrared Spectrum of Ice Ih and Ice Ic from 360 to
  50 cm$^{-1}$},'' {\em J. Chem. Phys.} {\bf 46}, 1271--1284  (1967).

\bibitem{jenniskens1995}
P.~{Jenniskens}, D.~F. {Blake}, M.~A. {Wilson}, {\em et~al.}, ``{High-Density
  Amorphous Ice, the Frost on Interstellar Grains},'' {\em Astrophys. J.} {\bf
  455}, 389  (1995).

\bibitem{hudgins1993}
D.~M. {Hudgins}, S.~A. {Sandford}, L.~J. {Allamandola}, {\em et~al.}, ``{Mid-
  and far-infrared spectroscopy of ices - Optical constants and integrated
  absorbances},'' {\em Astrophys. J. Suppl.} {\bf 86}, 713--870  (1993).

\bibitem{bertie1969}
J.~E. {Bertie}, H.~J. {Labb{\'e}}, and E.~{Whalley}, ``{Absorptivity of Ice I
  in the Range 4000--30\, cm$^{-1}$},'' {\em J. Chem. Phys.} {\bf 50},
  4501--4520  (1969).

\bibitem{moore1992}
M.~H. {Moore} and R.~L. {Hudson}, ``{Far-infrared spectral studies of phase
  changes in water ice induced by proton irradiation},'' {\em Astrophys. J.}
  {\bf 401}, 353--360  (1992).

\bibitem{smith1994}
R.~G. {Smith}, G.~{Robinson}, A.~R. {Hyland}, {\em et~al.}, ``{Molecular ices
  as temperature indicators for interstellar dust: the 44- and 62-{$\mu$}m
  lattice features of H$_{2}$O ice.},'' {\em Mon. Not. R. Astron. Soc.} {\bf
  271}, 481--489  (1994).

\bibitem{draine1985}
B.~T. {Draine}, ``{Tabulated optical properties of graphite and silicate
  grains},'' {\em Astrophys. J. Suppl.} {\bf 57}, 587--594  (1985).

\bibitem{zubko1996}
V.~G. {Zubko}, V.~{Mennella}, L.~{Colangeli}, {\em et~al.}, ``{Optical
  constants of cosmic carbon analogue grains - I. Simulation of clustering by a
  modified continuous distribution of ellipsoids},'' {\em Mon. Not. R. Astron.
  Soc.} {\bf 282}, 1321--1329  (1996).

\bibitem{warren2008}
S.~G. {Warren} and R.~E. {Brandt}, ``{Optical constants of ice from the
  ultraviolet to the microwave: A revised compilation},'' {\em Journal of
  Geophysical Research (Atmospheres)} {\bf 113}, D14220  (2008).

\bibitem{compiegne2011}
M.~{Compi{\`e}gne}, L.~{Verstraete}, A.~{Jones}, {\em et~al.}, ``{The global
  dust SED: tracing the nature and evolution of dust with DustEM},'' {\em
  Astron. Astrophys.} {\bf 525}, A103  (2011).

\bibitem{Mathis1983}
J.~S. {Mathis}, P.~G. {Mezger}, and N.~{Panagia}, ``{Interstellar radiation
  field and dust temperatures in the diffuse interstellar medium and in giant
  molecular clouds},'' {\em Astron. Astrophys.} {\bf 128}, 212--229  (1983).

\bibitem{Kamp2018}
I.~{Kamp}, A.~{Scheepstra}, M.~{Min}, {\em et~al.}, ``{Diagnostic value of
  far-IR water ice features in T Tauri disks},'' {\em Astron. Astrophys.} {\bf
  617}, A1  (2018).

\bibitem{Kamp2021}
I.~{Kamp}, M.~{Honda}, H.~{Nomura}, {\em et~al.}, ``{The formation of planetary
  systems with SPICA},'' {\em Publ. Astron. Soc. Aust.} {\bf 38}, e055  (2021).

\bibitem{lutz1996}
D.~{Lutz}, H.~{Feuchtgruber}, R.~{Genzel}, {\em et~al.}, ``{SWS observations of
  the Galactic center.},'' {\em Astron. Astrophys.} {\bf 315}, L269--L272
  (1996).

\bibitem{indebetouw2005}
R.~{Indebetouw}, J.~S. {Mathis}, B.~L. {Babler}, {\em et~al.}, ``{The
  Wavelength Dependence of Interstellar Extinction from 1.25 to 8.0 {$\mu$}m
  Using GLIMPSE Data},'' {\em Astrophys. J.} {\bf 619}, 931--938  (2005).

\bibitem{flaherty2007}
K.~M. {Flaherty}, J.~L. {Pipher}, S.~T. {Megeath}, {\em et~al.}, ``{Infrared
  Extinction toward Nearby Star-forming Regions},'' {\em Astrophys. J.} {\bf
  663}, 1069--1082  (2007).

\bibitem{wang2013}
S.~{Wang}, J.~{Gao}, B.~W. {Jiang}, {\em et~al.}, ``{The Mid-infrared
  Extinction Law and its Variation in the Coalsack Nebula},'' {\em Astrophys.
  J.} {\bf 773}, 30  (2013).

\bibitem{Declei2022}
M.~{Decleir}, K.~D. {Gordon}, J.~E. {Andrews}, {\em et~al.}, ``{SpeX
  Near-infrared Spectroscopic Extinction Curves in the Milky Way},'' {\em
  Astrophys. J.} {\bf 930}, 15  (2022).

\bibitem{Gordon2023}
K.~D. {Gordon}, G.~C. {Clayton}, M.~{Decleir}, {\em et~al.}, ``{One Relation
  for All Wavelengths: The Far-ultraviolet to Mid-infrared Milky Way
  Spectroscopic R(V)-dependent Dust Extinction Relationship},'' {\em Astrophys.
  J.} {\bf 950}, 86  (2023).

\bibitem{wang2014}
S.~{Wang}, A.~{Li}, and B.~W. {Jiang}, ``{Modeling the infrared interstellar
  extinction},'' {\em Planet. Sp. Sci.} {\bf 100}, 32--39  (2014).

\bibitem{boogert2011}
A.~C.~A. {Boogert}, T.~L. {Huard}, A.~M. {Cook}, {\em et~al.}, ``{Ice and Dust
  in the Quiescent Medium of Isolated Dense Cores},'' {\em Astrophys. J.} {\bf
  729}, 92  (2011).

\bibitem{steinacker2015}
J.~{Steinacker}, M.~{Andersen}, W.-F. {Thi}, {\em et~al.}, ``{Grain size limits
  derived from 3.6 {$\mu$}m and 4.5 {$\mu$}m coreshine},'' {\em Astron.
  Astrophys.} {\bf 582}, A70  (2015).

\bibitem{Ysard2016}
N.~{Ysard}, M.~{K{\"o}hler}, A.~{Jones}, {\em et~al.}, ``{Mantle formation,
  coagulation, and the origin of cloud/core shine. II. Comparison with
  observations},'' {\em Astron. Astrophys.} {\bf 588}, A44  (2016).

\bibitem{Saajasto2018}
M.~{Saajasto}, M.~{Juvela}, and J.~{Malinen}, ``{Near-infrared scattering as a
  dust diagnostic},'' {\em Astron. Astrophys.} {\bf 614}, A95  (2018).

\bibitem{boogert2015}
A.~C.~A. {Boogert}, P.~A. {Gerakines}, and D.~C.~B. {Whittet}, ``{Observations
  of the icy universe.},'' {\em Annu. Rev. Astron. Astrophys.} {\bf 53},
  541--581  (2015).

\bibitem{westphal2014}
A.~J. {Westphal}, R.~M. {Stroud}, H.~A. {Bechtel}, {\em et~al.}, ``{Evidence
  for interstellar origin of seven dust particles collected by the Stardust
  spacecraft},'' {\em Science} {\bf 345}, 786--791  (2014).

\bibitem{smith1989}
R.~G. {Smith}, K.~{Sellgren}, and A.~T. {Tokunaga}, ``{Absorption features in
  the 3 micron spectra of protostars},'' {\em Astrophys. J.} {\bf 344},
  413--426  (1989).

\bibitem{terada2012}
H.~{Terada} and A.~T. {Tokunaga}, ``{Discovery of Crystallized Water Ice in a
  Silhouette Disk in the M43 Region},'' {\em Astrophys. J.} {\bf 753}, 19
  (2012).

\bibitem{omont1990}
A.~{Omont}, S.~H. {Moseley}, T.~{Forveille}, {\em et~al.}, ``{Observations of
  40-70 micron bands of ice in IRAS 09371 + 1212 and other stars},'' {\em
  Astrophys. J. Lett.} {\bf 355}, L27--L30  (1990).

\bibitem{Sylvester1999}
R.~J. {Sylvester}, F.~{Kemper}, M.~J. {Barlow}, {\em et~al.}, ``{2.4-197 mu m
  spectroscopy of OH/IR stars: the IR characteristics of circumstellar dust in
  O-rich environments},'' {\em Astron. Astrophys.} {\bf 352}, 587--599  (1999).

\bibitem{molinari1999}
S.~{Molinari}, C.~{Ceccarelli}, G.~J. {White}, {\em et~al.}, ``{Detection of
  the 62 Micron Crystalline H$_{2}$O Ice Feature in Emission toward HH 7 with
  the Infrared Space Observatory Long-Wavelength Spectrometer},'' {\em
  Astrophys. J. Lett.} {\bf 521}, L71--L74  (1999).

\bibitem{dartois1998}
E.~{Dartois}, P.~{Cox}, P.~R. {Roelfsema}, {\em et~al.}, ``{Detection of the
  ``44\,$\mu$m'' band of water ice in absorption in combined ISO SWS-LWS
  spectra},'' {\em Astron. Astrophys.} {\bf 338}, L21--L24  (1998).

\bibitem{min2016}
M.~{Min}, J.~{Bouwman}, C.~{Dominik}, {\em et~al.}, ``{The abundance and
  thermal history of water ice in the disk surrounding HD 142527 from the DIGIT
  Herschel Key Program},'' {\em Astron. Astrophys.} {\bf 593}, A11  (2016).

\bibitem{McClure2012}
M.~K. {McClure}, P.~{Manoj}, N.~{Calvet}, {\em et~al.}, ``{Probing Dynamical
  Processes in the Planet-forming Region with Dust Mineralogy},'' {\em
  Astrophys. J. Lett.} {\bf 759}, L10  (2012).

\bibitem{McClure2015}
M.~K. {McClure}, C.~{Espaillat}, N.~{Calvet}, {\em et~al.}, ``{Detections of
  Trans-Neptunian Ice in Protoplanetary Disks},'' {\em Astrophys. J.} {\bf
  799}, 162  (2015).

\bibitem{jenniskens1996}
P.~{Jenniskens} and D.~F. {Blake}, ``{Crystallization of Amorphous Water Ice in
  the Solar System},'' {\em Astrophys. J.} {\bf 473}, 1104  (1996).

\bibitem{tachibana2017}
S.~{Tachibana}, A.~{Kouchi}, T.~{Hama}, {\em et~al.}, ``{Liquid-like behavior
  of UV-irradiated interstellar ice analog at low temperatures},'' {\em Science
  Advances} {\bf 3}, eaao2538  (2017).

\bibitem{tanaka2019}
K.~K. {Tanaka} and Y.~{Kimura}, ``{Theoretical analysis of crystallization by
  homogeneous nucleation of water droplets},'' {\em Physical Chemistry Chemical
  Physics (Incorporating Faraday Transactions)} {\bf 21}, 2410--2418  (2019).

\bibitem{Cuppen2024}
H.~M. {Cuppen}, H.~{Linnartz}, and S.~{Ioppolo}, ``{Laboratory and
  Computational Studies of Interstellar Ices},'' {\em Annu. Rev. Astron.
  Astrophys.} {\bf 62}, 243--286  (2024).

\bibitem{Cruz-Diaz2018}
G.~A. {Cruz-Diaz}, R.~{Mart{\'\i}n-Dom{\'e}nech}, E.~{Moreno}, {\em et~al.},
  ``{New measurements on water ice photodesorption and product formation under
  ultraviolet irradiation},'' {\em Mon. Not. R. Astron. Soc.} {\bf 474},
  3080--3089  (2018).

\bibitem{Oberg2009}
K.~I. {{\"O}berg}, H.~{Linnartz}, R.~{Visser}, {\em et~al.}, ``{Photodesorption
  of Ices. II. H$_{2}$O and D$_{2}$O},'' {\em Astrophys. J.} {\bf 693},
  1209--1218  (2009).

\bibitem{Draine1990}
B.~T. {Draine}, ``{Mass determinations from far-infrared observations},'' in
  {\em The Interstellar Medium in Galaxies},  H.~A. {Thronson}, Jr. and J.~M.
  {Shull}, Eds., {\em Astrophysics and Space Science Library} {\bf 161},
  483--492  (1990).

\bibitem{draine1983}
B.~T. {Draine}, W.~G. {Roberge}, and A.~{Dalgarno}, ``{Magnetohydrodynamic
  shock waves in molecular clouds},'' {\em Astrophys. J.} {\bf 264}, 485--507
  (1983).

\bibitem{hollenbach1997}
D.~J. {Hollenbach} and A.~G.~G.~M. {Tielens}, ``{Dense Photodissociation
  Regions (PDRs)},'' {\em Annu. Rev. Astron. Astrophys.} {\bf 35}, 179--216
  (1997).

\bibitem{lesaffre2013}
P.~{Lesaffre}, G.~{Pineau des For{\^e}ts}, B.~{Godard}, {\em et~al.},
  ``{Low-velocity shocks: signatures of turbulent dissipation in diffuse
  irradiated gas},'' {\em Astron. Astrophys.} {\bf 550}, A106  (2013).

\end{thebibliography}
\bibliographystyle{spiejour}   % makes bibtex use spiejour.bst

%%%%% Biographies of authors %%%%%

\vspace{2ex}\noindent\textbf{Takashi Onaka} is a professor emeritus at the University of Tokyo. He received his BS, MS, and PhD degrees in astronomy from the University of Tokyo in 1975, 1977, and 1980, respectively.  He is the author of more than 300 journal papers. He was the PI of Infrared Camera (IRC) on board the Japanese infrared satellite AKARI.  His current research interests include interstellar dust grains and infrared observations.

\vspace{2ex}\noindent\textbf{Itsuki Sakon} is an associate professor at the University of Tokyo.  He received his BS, MS, and PhD degrees in astronomy from the University of Tokyo in 2003, 2005, and 2008, respectively. His current research interests include near- and mid-infrared astronomical instrumentation development, infrared observational astronomy, laboratory astrophysics, and astrochemistry of organic matter in space. 

\vspace{2ex}\noindent\textbf{Takashi Shimonishi} is an associate professor at Niigata University.  He received his BS degree from Tohoku University in 2007, and MS and PhD degrees in astronomy from the University of Tokyo in 2009 and 2012, respectively.  His current research interests include interstellar chemistry in low-metallicity environments and observations of interstellar molecules.

\vspace{2ex}\noindent\textbf{Mitsuhiko Honda} is an associate professor at Okayama University of Science.  He received his BS, MS and PhD degrees from the University of Tokyo in 2000, 2002 and 2005, respectively. His current research interests include the material evolution through planet formation and mid-infrared observations.

\vspace{1ex}
\noindent Photographs of the authors are not available.

\listoffigures
\listoftables

\end{spacing}
\end{document}